# Adaptive SpikeDeep-Classifier: Self-organizing and self-supervised machine learning algorithm for online spike sorting

Muhammad Saif-ur-Rehman[1]*, Omair Ali[2,3]*, Christian Klaes[2] and Ioannis Iossifidis[1]

[1] Department of Computer Science, Ruhr-West University of Applied Science, Mülheim an der Ruhr, Germany; [2] Faculty of Medicine, Department of Neurosurgery, University hospital Knappschaftskrankenhaus Bochum GmbH, Germany, [3] Department of Electrical Engineering and Information Technology, Ruhr-University Bochum

## Abstract

***Objective.*** Invasive brain-computer interface (BCI) research is progressing towards the realization of the motor skills rehabilitation of severely disabled patients in the real world. The size of invasively implanted microelectrode arrays and the selection of an efficient online spike sorting algorithm are two key factors that play pivotal roles in the successful decoding of the user intentions. Recently, a very small but dense microelectrode array with 3072 channels was developed and implanted to decode the intention of the user. The process of spike sorting includes the selection of channels that record the spike activity (SA) and determines the SA of different sources (neurons), on selected channels individually. The neural data recorded with dense microelectrode arrays is time-varying and often contaminated with non-stationary noise. Unfortunately, currently available state-of-the-art spike sorting algorithms are incapable of handling the massively increasing amount of time-varying data resulting from the dense microelectrode arrays, which makes the spike sorting one of the fragile components of the online BCI decoding framework. ***Approach.*** This study proposed an adaptive and self-organized algorithm for online spike sorting, named as Adaptive SpikeDeep-Classifier (Ada-SpikeDeepClassifier). Our algorithm uses SpikeDeeptector for the channel selection, an adaptive background activity rejector (Ada-BAR) for discarding the background events, and an adaptive spike classifier (Ada-Spike classifier) for classifying the SA of different neural units. By concatenating SpikeDeeptector, Ada-BAR and Ada-Spike classifier, the process of spike sorting is accomplished. ***Results.*** The proposed algorithm is evaluated on two different categories of data: a human dataset recorded in our lab, and a publicly available simulated dataset to avoid subjective biases and labeling errors. The proposed Ada-SpikeDeepClassifier outperformed our previously published SpikeDeep-Classifier and eight other spike sorting algorithms. ***Significance.*** To the best of our knowledge, the proposed algorithm is the first spike sorting algorithm that automatically learns the abrupt changes in the distribution of noise and SA. The proposed algorithm is artificial neural network-based, which makes it an ideal candidate for its hardware implementation on neuromorphic chips that is also suitable for wearable invasive BCI.

# 1. Introduction

The human brain is one of the most complex systems in this observable universe, with approximately 100 billion information processing units called neurons and one trillion synapses per cubic centimeter of the cortex (Herculano-Houzel, 2009). Neurons propagate information via rapidly changing membrane potentials called action potentials. These action potentials are represented as sudden, propagatory and transitory changes in the resting state membrane potentials, thus, also referred to as spike activity (SA).

In electrophysiology, invasively implanted microelectrode arrays are the primary source of recording the SA of the neurons in the target brain areas. It is important to record the SA of a large population of neurons to understand the relationship between neuronal activity and behavioral functionality e.g. relationship between imagining a hand movement and the SA of corresponding brain areas (Aflalo et al., 2015), (Ali et al., 2021), (Ali et al., 2022), and (Klaes et al., 2015a). Henceforth, instead of a single microelectrode array, often multiple microelectrode arrays with several hundred channels (Harris et al., 2016) are implanted simultaneously.

Even though high-density microelectrode arrays enable recording the SA of several hundred neurons, it is not trivial to extract the SA of individual neurons from the raw data using a spike sorting algorithm due to the following reasons:
- A considerable fraction of the channels of the implanted microelectrode arrays do not record the SA but the artifacts, which are composed of non-stationary muscle artifacts, background activity generated by the neuronal activity too far away from the tip of the microelectrode and the noise generated from surrounding electrical equipment.
- The remaining electrodes which record the SA are also contaminated with background activity (BA).

Therefore, an efficient and reliable spike sorting algorithm should:
- Efficiently and accurately determine and discard the channels that exclusively record artifacts.
- Remove the background activity (BA) from the channels which record SA but are still contaminated with some BA.
- Accurately estimate the number of distinct neural units of each channel and determine the corresponding SA of these neural units.

The above explained process of spike sorting can be accomplished in three different modes: manual, semi-automatic and fully automatic. Manual and semi-automatic algorithms(Abeles and Goldstein, 1977); (Lewicki, 1998); (Gibson et al., 2012) require human curation at one or more stages of the spike sorting pipeline. Therefore, the process becomes labor-intensive and time-consuming. Therefore, these methods cannot process the ample amount of data being recorded with high-density microelectrode arrays in real time. Besides that, the spike sorting results of manual and semi-automatic algorithms rely solely on the expertise of a curator. Even an expert neuroscientist with several years of spike sorting experience, may come up with slightly different conclusions about the same data on two different occasions.

The automatic spike sorting algorithms(Bongard et al., 2014; Carlson et al., 2014; Chung et al., 2017; Hu et al., 2022; Pachitariu et al., 2016; Spacek et al., 2009; Takekawa et al., 2012; Yger et al., 2018) attempt to address the flaws of manual and semi-automatic spike sorting algorithms by completely automating the spike sorting pipeline. However, the neural data is non-stationary and time-varying. By non-stationary, we mean that the implanted microelectrode arrays may slightly move during two consecutive recording sessions. Consequently, implanted microelectrodes may start reading the data from new units or a few of the microelectrodes start reading external artifacts, which results in a change in the distribution of data that consequently invalidates the optimized values of the parameters of an already trained model. As a result, the fine-tuning of the parameters is required to acquire the optimal performance of the online spike sorting algorithms for the new distribution of the data. However, the process of tuning of parameters is required for each recording session, which is not a desirable solution. This lack in generalization of the available spike sorting algorithms is due to the non-stationary behavior and time-varying characteristics of BA and SA because of the small perturbations of the implanted microelectrodes.

In (Saif-ur-Rehman et al., 2019), we proposed a generalized solution (SpikeDeeptector) for the selection of the channels which record the neural activity and discard the channels which exclusively record artifacts. This is the first step of any spike sorting algorithm. The proposed SpikeDeeptector algorithm was trained on the versatile training dataset and enabled contextual learning by constructing the feature vectors by concatenating the batch of the events together. We extended the SpikeDeeptector algorithm and proposed a complete offline spike sorting pipeline called SpikeDeep-Classifier in (Saif-ur-Rehman et al., 2020). The proposed SpikeDeep-Classifier is an amalgamation of supervised and unsupervised machine learning algorithms. Even though the performance of the SpikeDeep-Classifier on the majority channels is quite impressive, however on few channels, where there is a high overlap between SA and BA events, the performance degrades. Another major drawback is the lack of online spike sorting ability of the SpikeDeep-Classifier algorithm. In this study, we updated the SpikeDeep-Classifier algorithm and proposed a self-supervised, self-organizing and online spike sorting algorithm called Ada-SpikeDeep-Classifier. The proposed algorithm automatically adapts to the changes in the distribution of the data by generating pseudo labels and fine-tunes itself. The proposed algorithm generally outperforms the SpikeDeep-Classifier, more specifically, on the data of the distinct classes located closer to the decision boundaries.

## 2. Materials & Methods

### 2.1 Approval

For this study, we used data from two tetraplegic patients who were implanted with two Utah arrays (Blackrock Microsystems, Salt Lake City, UT). These patients were recruited for two brain-computer interface studies (Aflalo et al., 2015; Klaes et al., 2015a), which were approved by the institutional review boards at the California Institute of Technology (Pasadena, CA), Rancho Los Amigos National Rehabilitation Center (Downey, CA), and the University of Southern California (USC) (Los Angeles, CA). Further approval information is available in (Aflalo et al., 2015; Klaes et al., 2015a).

### 2.2 Demographic & Implantation details

Two randomly selected recording sessions of tetraplegic patients were considered for the evaluation of the proposed Ada-SpikeDeepClassifier algorithm. These patients were implanted with two Utah arrays on the posterior parietal cortex (PPC). Each Utah array contains 100 electrodes arranged in a 10X10 grid. During manufacturing, 4 corner electrodes of the Utah array remain unconnected, thus 96 electrodes record the data. Utah array electrodes are 1.0 to 1.5 mm long and presumably can record the SA from cortical layer 5. The placement of the electrode arrays was based on functional magnetic resonance imaging (fMRI) tasks conducted prior to implantation. Detailed information about the placement of the microelectrode arrays is available in (Klaes et al., 2015b) and (Aflalo et al., 2015).

Table 1: Implantation and demographic details

| Subject Id | Sex | Age (year) | Place of implantation | Number of recording sessions | Number of electrodes |
|---|---|---|---|---|---|
| U1 | Male | 32 | PPC | 1 | 96 (Utah array) |
| U2 | Male | 63 | PPC | 1 | 96 (Utah array) |

### 2.3 Data Recording and Preprocessing

The data was recorded (digitized and preprocessed) by a neural signal processor (NSP) (Blackrock Microsystems, Salt Lake City, UT). In this study, we used an end-to-end machine learning solution (Glasmachers, 2017; Lecun et al., 1998) for spike sorting problem, therefore, minimal preprocessing is applied and only recording hardware performs the preprocessing, which involves extraction of events (SA candidates) from the raw data.

Amplitude is one of the most prominent features which characterize the shape of detected events (SA candidates). Therefore, amplitude thresholding (Lewicki, 1998) is widely applied to extract events. In this study, the events are extracted by employing dynamic thresholding across amplitude.

Suppose $X = (x_1, x_2, .......... x_n)$ is a filtered signal (high-pass filtered, full-bandwidth signal with cutoff frequency 250 Hz). The threshold setting for spike candidate extraction was set at -4.5 times root-mean-square (rms) of the input signal, as shown in the equation. The given threshold setting (scalar multiplication factor) was used because it was previously used in online decoding studies (Aflalo et al., 2015; Klaes et al., 2015b).

$$\text{Threshold} = -4.5 \sqrt{\frac{1}{T} \sum_{x_i=1}^{T} x_i^2}$$

### 2.4 Data labeling

For this study, we used the labeled data for the evaluation of the proposed Ada-SpikeDeepClassifier algorithm. The extracted thresholded events are either labeled as spike activity (SA) or background activity (BA). The events which represent the action potentials are labeled as SA. Contrarily, the events that represent technical artifacts or neural activity too far away from the tip of the electrode are labeled as BA. Further, the SA events may represent the activity of more than neurons, and further labeled into (unit 1, unit 2, … unit n) subcategories. The labeling of data is done using a gaussian mixture model (GMM) based in-house developed semi-automatic offline spike sorting framework. Comprehensive explanation of data labeling process is available in Data labeling section of (Saif-ur-Rehman et al., 2019).

### 2.5 Ada-SpikeDeepClassifier

This study presents a self-organizing, adaptive, semi-online and fully automatic spike sorting algorithm called Ada-SpikeDeepClassifier. The architecture of Ada-SpikeDeepClassifier which accomplishes spike sorting in three stages is shown in **Figure 1**. The proposed algorithm first discards the channels that exclusively record background activity in combination with external artifacts using **SpikeDeeptector** algorithm (Saif-ur-Rehman et al., 2019). In the next stage, thresholded BA events of the remaining channels were discarded using Adaptive **background active rejector (Ada-BAR)**. In the final stage, SA which may represent the activity of more than one neuron are then categorized into sub-classes of SA using a **self-organizing sorter**. More precisely, the self-organizing sorter determines the number of distinct units and their activity on each channel that records the neural data.

Comprehensive explanation of each processing element of Ada-SpikeDeepClassifier is available in the following subsections.

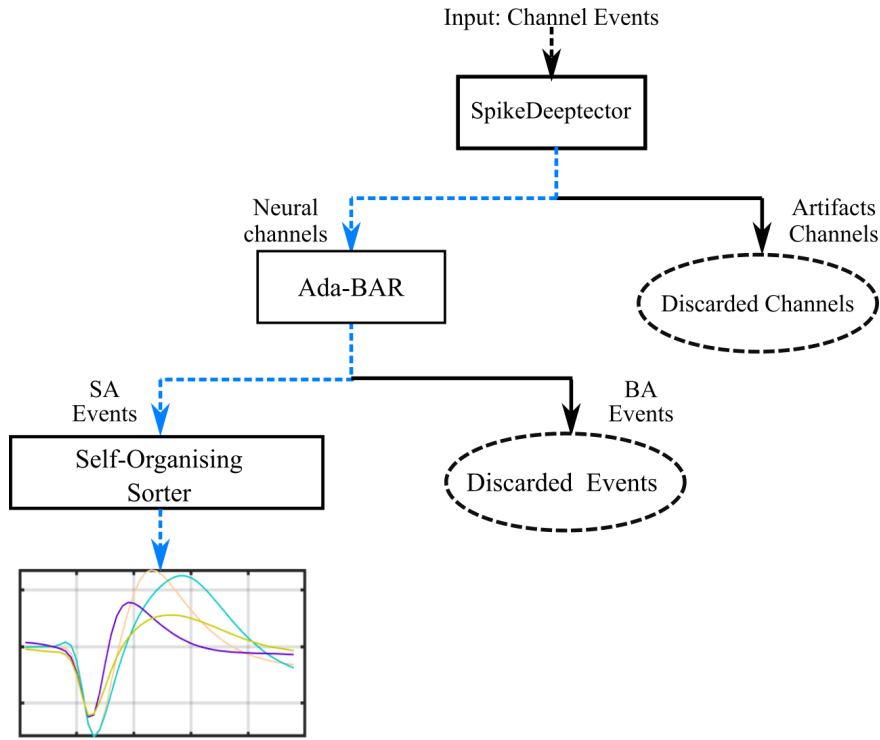

**Figure 1**: *Ada-SpikeDeepClassifier three processing stages*. In the first stage, SpikeDeeptector algorithm selects the channels recording neural data. In the second stage, the events of neural channels are fed to Ada-BAR, which classifies the events as SA or BA. In the third stage, SA events are then further classified into individual neural units using self-organizing sorter.

### 2.5.1 SpikeDeeptector

SpikeDeeptector algorithm is the first processing stage of Ada-SpikeDeepClassifier. SpikeDeeptector provides a generalized solution to determine the channels that record the neural data. By generalized solution, we mean that pre-trained model of SpikeDeeptector without any further fine-tuning can be used on new unseen data to determine the channels that record neural data. In (Saif-ur-Rehman et al., 2019), we showed for several recording sessions of human patients and non-human primates that the SpikeDeeptector successfully discriminates between the channels that record neural data and the channels that exclusively record artifacts.

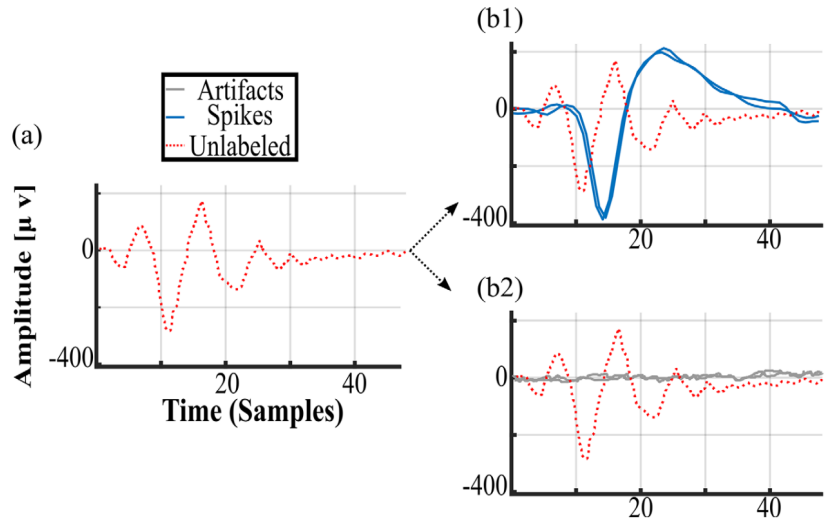

SpikeDeeptector discriminates by enabling contextual learning through the novel way of feature vector extraction. We embed the context in the feature vectors by concatenating the batch of events. If one or more of the concatenated events represent SA the feature vector was labeled as '*Spike*', contrarily, if all the concatenated events represent 'BA' the feature vector was labeled '*Artifact*'. SpikeDeeptector follows standard architecture of convolutional neural networks (Krizhevsky et al., 2017; Lecun et al., 1998). The regularized (L2-regularization) cross entropy cost function was minimized using mini-batch gradient descent with momentum. Further, dropout regularization was used to avoid overfitting. A comprehensive explanation of SpikeDeeptector is available in "SpikeDeeptector algorithm" section of (Saif-ur-Rehman et al., 2019).

Six recording sessions of a tetraplegic patient implanted with Utah arrays were used for training SpikeDeeptector model. Then the trained model of SpikeDeeptector was evaluated on 130 test recording sessions. Evaluation performance of generalization of SpikeDeeptector is available in the Result section of (Saif-ur-Rehman et al., 2019).

### 2.5.2 Ada-BAR

**Background activity rejector (BAR):** BAR processes a 48-sampled event through 3 convolutional layers, 2 max-pooling layers, a fully connected layer and finally a softmax layer for the classification of a given event as spike activity (SA) or background activity (BA). A detailed explanation of the BAR algorithm and the process of tuning hyperparameters is available in the Background *activity rejector (BAR)* section of (Saif-ur-Rehman et al., 2020).

A robust and highly generalized BAR model was learned by training on the multiple recording sessions of two species, five subjects, six brain areas, three different types of microelectrodes and two recording systems. Exact contribution of each subject in the training dataset is available in section *BAR data distribution for training and validation* of (Saif-ur-Rehman et al., 2020).

**Compromised performance of BAR closer to decision boundary:** Even though, BAR works quite well on the majority data points of a channel, as shown in **Figure 2**. However, on few channels where the data points of two distinct classes were overlapped (see **Figure 2b)** performance of BAR was found to be suboptimal. Therefore, to address this issue, an adaptive version of BAR is proposed in this study.

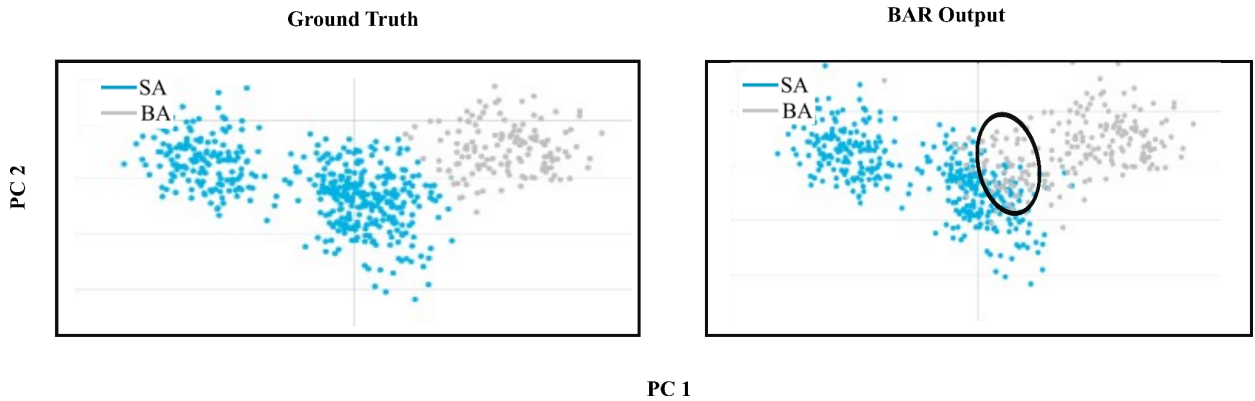

**Figure 2**: *Evaluation of BAR on the data points closer to decision boundaries. Performance BAR on the overlapped data points of distinct classes compromised.*

**Ada-BAR:** In this study, a self-supervised learning algorithm is proposed to address the suboptimal performance of BAR on the data, which lies in the vicinity of decision boundaries. We named the proposed algorithm Ada-BAR due to its adaptive nature to the new unseen data.

The working principles of Ada-BAR are elaborated in **Figure 3.** The flow of the data takes place from left to right side as shown in **Figure 3**. BAR predicts the events of a channel as SA or BA.

Firstly, unlabeled events of a channel are provided to BAR, which predicts the labels of these events. The same data is also provided to K-Means clustering algorithms, initialized with the number of clusters much higher than the expected number of clusters in the provided data. Precisely, K-Means clustering algorithms were asked to find 7 clusters in the given data. Next stage determines the consensus between Global BAR and K-Means clustering algorithms. Here, consistency of the BAR predictions was evaluated on the data of each cluster, individually. If at least 70% predictions of the Global BAR on the data of a cluster belong to one particular class (SA or BA). Then, all the data of that particular cluster is assigned to that particular class and used as pseudo labels. Otherwise, the data is not further used. In the last stage, only 25% of the pseudo labeled data is used to fine-tune the Global BAR to adapt the change in the distribution of the data. The fine-tuned model is called the Ada-BAR.

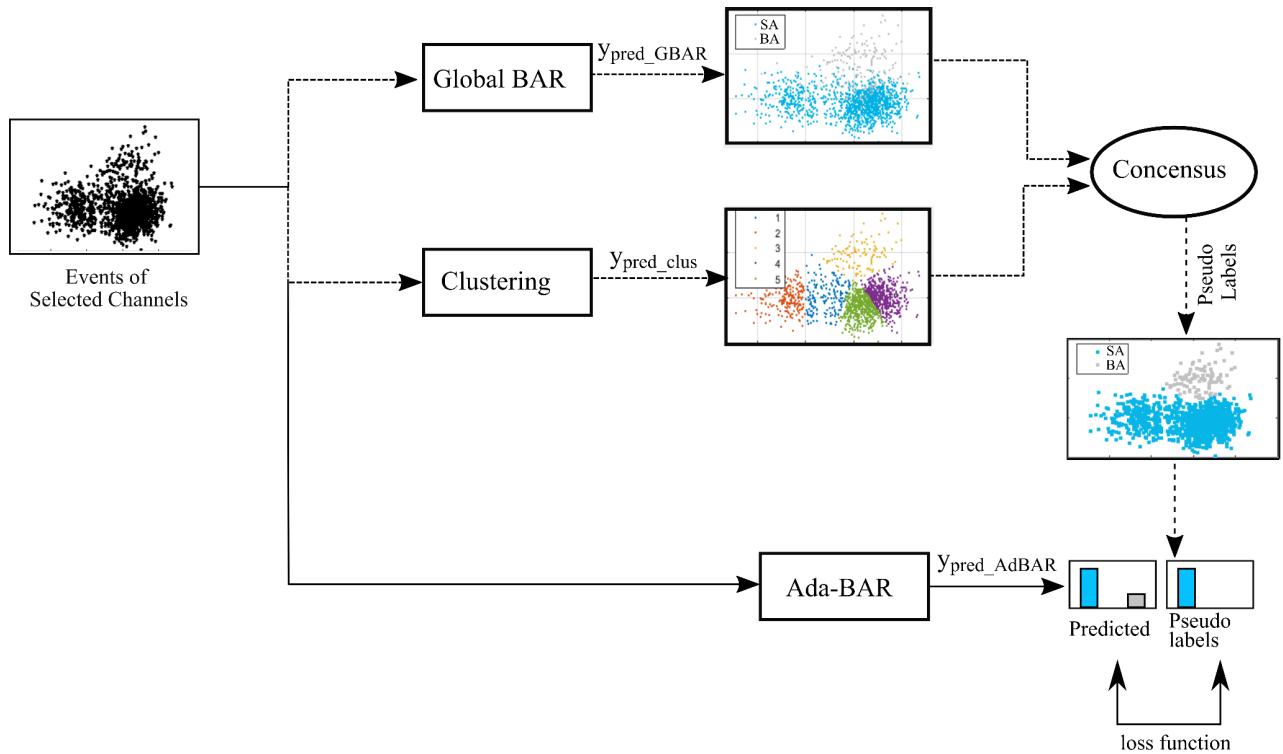

**Figure 3**: *Ada-BAR working principles: pretrained model of BAR is fine-tuned in a self-supervised way.* Pseudo labels are produced by the consensus of pretrained BAR predictions and a clustering algorithm predictions (K-means) with the total number of clusters higher than the actual existing clusters in the presented data. Only 25% of the pseudo labels were used to fine-tune the pretrained BAR model, the resultant model is called Ada-BAR.

### 2.5.3 Self-Organizing Sorter

Filtered data resulting from SpikeDeeptector in conjunction with Ada-BAR is provided to the self-organizing sorter to identify the total number of distinct neural units with their corresponding activity of a channel. This process takes place in two stages, as shown in **Figure 4 (a, b).**

In the first stage (see **Figure 4 (a)**), a convolutional neural network is trained on the labeled SA events of a channel with three distinct neural units. CNN processes the given inputs through 3 convolutional layers, 2 max-pooling layers, a fully connected layer and finally a softmax layer for the classification of a given spike activity (SA). A detailed description of the architecture and optimization process of tuning the parameters and hyperparameters is discussed in **Supplementary materials**.

The second stage of the self-organizing sorter consists of self-supervised and self-organizing architecture of convolutional neural network as shown in **Figure 4 (b)**. Here, unlabeled SA events are processed through pre-trained convolutional layers, K-mean clustering with predefined maximum number of clusters, and CAOM algorithm to generate pseudo labels. The number of neurons of the last dense layer (Softmax layer) is equal to distinct neural units recorded by that particular channel, however, the number of distinct neural units on different channels are different. Therefore, introduction of self-organizing attribute to the CNN architecture (**Figure 4 (b)**)  to determine the number of neurons of the last layer was required. We determined the number of neuron on a channel using CAOM algorithm, which was developed in our previous study  (Saif-ur-Rehman et al., 2020) to automatize the clustering process. After the generation of Pseudo labels, the complete architecture of the convolutional neural network defined in **Figure 4 (b)** was fine-tuned using 40% of the data and evaluated on the remaining data.

Further explanation and working principles K-means and CAOM algorithm is available in **Clustering Method** and **Cluster Accept or Merge Algorithm** section of the (Saif-ur-Rehman et al., 2020).

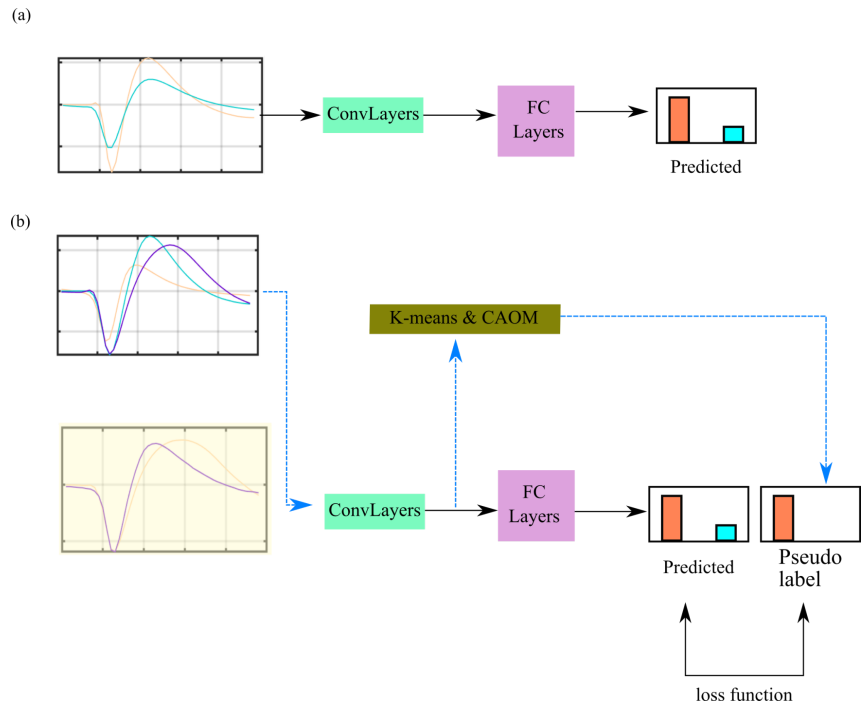

**Figure 4**: *Self-Organizing Sorter working principles*: *(a) Step1: A conv net is trained from the scratch on the labeled data of a channel with two distinct neural units. (b) Step 2: Convolutional layers of trained conv net (resulted in step 1) and K-means in conjunction with cluster accept or merge (CAOM) algorithm are used to generate the pseudo labels as well as to determine the number of neurons of the last dense layer.*

# 3. Results

### 3.1 Evaluation Metrics

We reported the classification accuracy of SpikeDeeptector and Ada-BAR. Furthermore, to enhance the transparency of the reported results, Recall is used as another evaluation metric for Ada-BAR. Mathematical representation of both the accuracy and recall is shown in **Equation 1** and **Equation 2**.

$$Accuracy = \left(\frac{Number\ of\ correct\ predictions}{Total\ number\ of\ examples}\right) \times 100 \quad (1)$$

$$Recall = \left(\frac{True\ positives}{True\ postives + False\ negatives}\right) \times 100 \quad (2)$$

In addition, accuracy and RandIndex are reported to evaluate the performance of Ada-Sorter.

### 3.2 SpikeDeeptector

### 3.2 Ada-BAR

Pre-trained models of BAR and self-supervised Ada-BAR were evaluated on the data of two recording sessions of two human patients implanted with Utah arrays. Ada-BAR yields improvement compared to its predecessor "BAR" on both the recording sessions, as shown in confusion matrices (see **Table 2**).

In recording session 1, Ada-BAR provides classification accuracy of 93.0% which is 6.4% higher than BAR. Correct classification of the data that represents SA is very crucial for spike sorting and invasive BCI decoding applications. Ada-BAR results in 94.7% classification accuracy on the SA, which is 7.3% higher than BAR (see **Table 2**). As a result, chances to discard meaningful data is reduced.

Similarly, Ada-BAR outperforms BAR on the data of the second recording session. Here, Ada-BAR yields 91.2% classification accuracy and BAR achieves 85.7% overall classification accuracy as shown in **Table 2**. Classification accuracy of Ada-BAR performs on SA data is accuracy 92.2% and the classification of BAR SA data is 85.9%.

**Table 2** provides a quantitative comparison of evaluation performance of BAR and Ada-BAR, where Ada-BAR yields better performance. However, it is important to inspect the performance of both the models with visualization, which is shown in **Figure 5**. The performance of BAR and Ada-BAR is comparable on the data points of both classes, which are far from each other. However, the most critical area is near to decision boundaries, where it is quite evident that Ada-BAR comprehensively outperforms BAR (see **Figure 5**). In addition, a quantitative comparison of BAR and Ada-BAR is also shown in **Figure 5,** where only 18 data points of SA were misclassified by Ada-BAR. In contrast, BAR misclassified 80 data points, which represent SA.

**Table 2**: *Confusion matrices:* Evaluation performance of BAR and Ada-BAR on two recording sessions of tetraplegic human patients implanted with Utah arrays. Ada-BAR yields improvements in classification accuracy, precision and Recall on both the recording session.

**Session: 1**

| | BAR | | | | Ada-BAR | | |
|---|---|---|---|---|---|---|---|
| SA | 44156 / 81.3% | 877 / 1.6% | 98.1% | SA | 47868 / 88.2% | 1136 / 2.1% | 97.7% |
| BA | 6385 / 11.8% | 2871 / 5.3% | 31.0% | BA | 2673 / 4.9% | 2612 / 4.8% | 49.4% |
| | 87.4% | 76.6% | **86.6%** | | 94.7% | 69.7% | **93.0%** |
| | SA | BA | | | SA | BA | |

**Session: 2**

| | BAR | | | | Ada-BAR | | |
|---|---|---|---|---|---|---|---|
| SA | 15006 / 71.7% | 502 / 2.4% | 96.8% | SA | 16096 / 76.9% | 466 / 2.2% | 97.2% |
| BA | 2459 / 11.7% | 2967 / 14.2% | 54.7% | BA | 1369 / 6.5% | 3003 / 14.3% | 68.7% |
| | 85.9% | 85.5% | **85.9%** | | 92.2% | 86.6% | **91.2%** |
| | SA | BA | | | SA | BA | |

Predicted Labels (y-axis) — True Labels (x-axis)

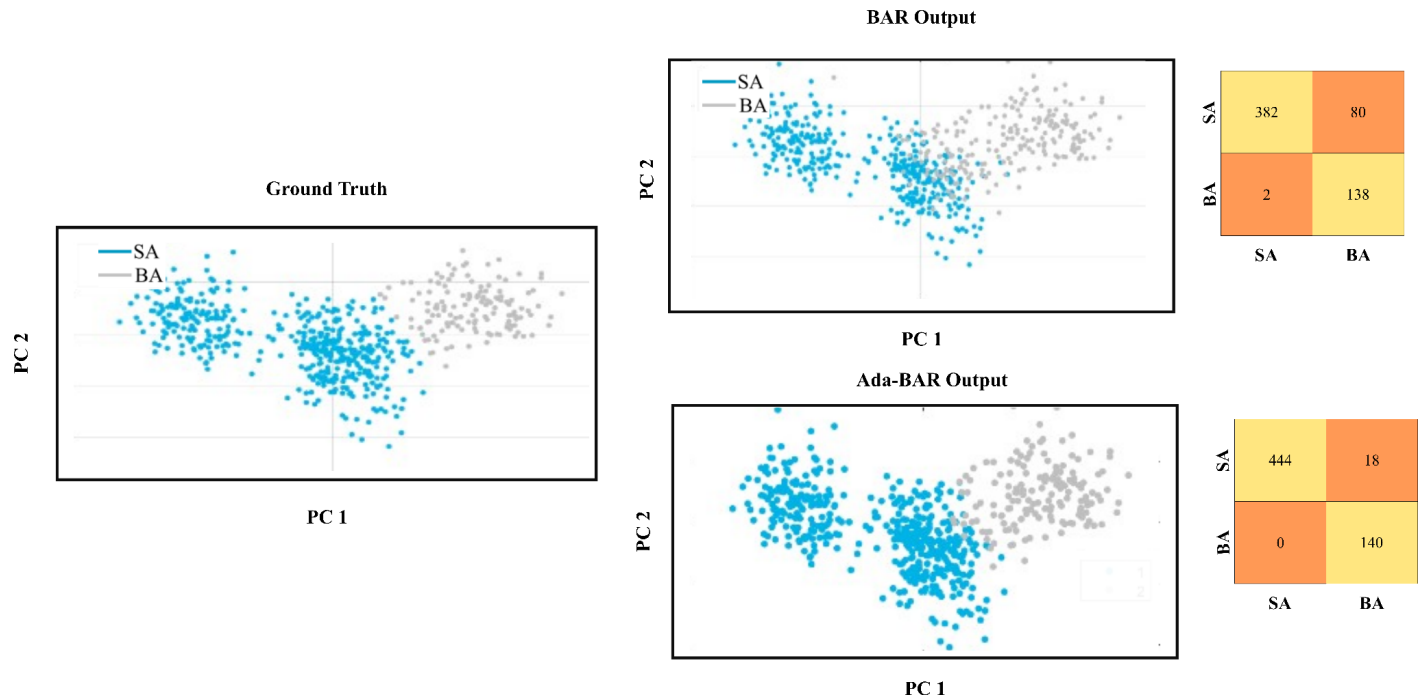

**Figure 5**: **Performance comparison of BAR and Ada-BAR on a channel with overlapping data points of SA and BA**: Ada-BAR managed to perform better than BAR, specifically closer to the decision boundary.

### 3.3 Self-organizing Sorter

**Table 3** and **Table 4** provide an overview of performance comparison of two algorithms employed to classify the SA of a channel into one or more neural units. K-means clustering algorithm which was conjugated with cluster accept or merge (CAOM) for automatic spike sorting was provided with the SA predicted by BAR. Contrarily, self-organizing sorter was employed on the SA predicted by Ada-BAR.

In both the recording sessions, out of 192 channels only 40 channels recorded SA (see **Table3 & Table 4**), from which 26 channels recorded one single unit, 12 channels recorded two single units and only 2 channels were able to record three single units. K-means clustering in combination with CAOM correctly predicted the number of neural units on 35 channels. In addition, rand index and classification accuracy were used to assess the quality of the prediction of the employed algorithm. The Rand index provides a measure of similarity between two data clustering methods. The value of the Rand index is between 0 and 1. 1 means that both clustering (ground truth, predicted) methods produced the exact same results, and 0 indicates that two data clustering methods completely disagree with each other.

For the recording session from 2014 the achieved mean rand index of self-organizing sorter is more than 0.84 for any number of neural units on a channel (see Table 3). Similarly, the achieved mean accuracy for any number of units is more than 89%. Here, the mean rand index for K-means + CAOM is 0.79 and the accuracy is 84.8%.

**Table 3**: *Performance comparison of SA classification*: Evaluation performance of K-means + CAOM and self-organizing sorter on a recording session recorded in 2014 of tetraplegic human patients implanted with Utah arrays.

| Number of Units | No. of channels (True) | No. of Channels (Pred.) (Correct, wrong) | Rand Index (K-means + CAOM) | Accuracy (K-means + CAOM) | Rand Index (Self-organizing sorter) | Accuracy (Self-organizing sorter) |
|---|---|---|---|---|---|---|
| 3 | 2 | (2,1) | 0.80 + 0.07 | 85.9 ± 4.0 | 0.86 ± 0.08 | 91 ± 4.9 |
| 2 | 7 | (5,1) | 0.78 + 0.06 | 82.1 ± 3.5 | 0.83 ± 0.08 | 87.7 ± 7.1 |
| 1 | 4 | (3,1) | 0.79 + 0.15 | 86.4 ± 7.5 | 0.85 ± 0.18 | 90.2 ± 9 |

For the recording session from 2013 the achieved mean rand index of self-organizing sorter is more than 0.90 for any number of neural units on a channel (see Table 3). Similarly, the achieved mean accuracy for any number of units is more than 91%. Here, the mean rand index for K-means + CAOM is 0.85 and the accuracy is 87.7%.

**Table 4**: *Performance comparison of SA classification*: Evaluation performance of K-means + CAOM and self-organizing sorter on a recording session recorded in 2013 of tetraplegic human patients implanted with Utah arrays.

| Number of Units | No. of channels (True) | No. of Channels (Pred.) (Correct, wrong) | Rand Index (K-means + CAOM) | Accuracy (K-means + CAOM) | Rand Index (Self-organizing sorter) | Accuracy (Self-organizing sorter) |
|---|---|---|---|---|---|---|
| 3 | 0 | (0,0) | NA | NA | NA | NA |
| 2 | 5 | (5, 1) | 0.89 ± 0.04 | 88.3 ± 9 | 0.94 ± 0.06 | 92.74 ± 9.8 |
| 1 | 22 | (20,1) | 0.81 ± 0.14 | 87.1 ± 10.9 | 0.86 ± 0.16 | 90.5 ± 13.45 |

## 4. Discussion

In this study, we proposed an updated and improved version of SpikeDeep-Classifier algorithm called Adaptive-SpikeDeepClassifier. SpikeDeep-Classifier. In our previous study (Saif-ur-Rehman et al., 2020), we showed that the trained model of SpikeDeep-Classfiier enabled spike sorting on the data of multiple species, multiple subjects who were implanted with different types of microeletroded (Utah array, Single microelectrode) on different brain areas (posterior parietal cortex and anterior hippocampus). SpikeDeep-Classifier completes the process of spike sorting by processing the raw data in four stages; SpikeDeeptector for the selection of the channels recording neural data, background activity rejector (BAR) for discarding BA from the selected selected channels, principal component analysis for dimensionality reduction of the resultant SA and finally K-means in conjunction cluster accept or merge (CAOM) algorithm for automatic identification of total distinct neural units along with their SA.

Even though, BAR works quite well on the majority data points of a channel, as shown in **Figure 2**. However, performance of BAR was found to be suboptimal on few data points of distinct classes which are quite close to each other (see **Figure 2b)**. We deployed one trained model of BAR without any further retraining or fine-tuning for the separation of BA and SA for all the used data. Therefore, it is possible that its performance is not equally good on all the data points of all the channels, especially on the data points closer to decision boundaries. We also do not want to manually label the data for fine-tuning the trained model of BAR on each channel separately. Therefore, in this study we developed an adaptive BAR, which works on the principles of self-supervised machine learning paradigm and automatically fine-tunes itself on self-generated pseudo labels of each individual channel, resulting in improvements of classification accuracy. Confusion matrices shown in Table 2 provide a comprehensive quantitative comparison of BAR and Ada-BAR, where Ada-BAR yields 6.4% and 5.3% improvement in classification accuracy. A visual performance comparison of both algorithms is shown in **Figure 5**, where it is quite clear that Ada-BAR performs much better on the data points quite close to the decision boundaries of two classes.

SpikeDeep-Classifier uses PCA and K-means in conjunction with CAOM for the further classification of SA data, which provides an offline solution for spike sorting problem. In this study, we developed a semi-online spike sorting using self-organizing sorter, which is also a self-supervised learning algorithm. Self-organizing sorter generates pseudo labels using K-means and CAOM and then fine-tune the pretrained model of convolutional neural network for final classification. After fine tuning, self-organizing sorter can be used in online mode.

Adaptive SpikeDeep-Classifier outperformed its counterpart SpikeDeep-Classifier on the recording sessions of tetraplegics implanted with Utah arrays. In addition, we compared Ada-SpikeDeepClassifier with eight existing spike sorting algorithms (Nguyen et al., 2015) as shown in **Table 5**. We used publicly available simulated dataset (Quiroga et al., 2004), which is actively used as a benchmark dataset for the comparison of spike sorting algorithms. The dataset is labeled and contains four examples with four noise levels (0.05, 0.1, 0.15, 0.2).

The spike sorting algorithms presented in (Nguyen et al., 2015) are automatic and accomplish the process of spike sorting in two steps, which are feature extraction, and automatic clustering. For feature extraction wavelet transform (WT) and diffusion maps (DM) were used and then the activity of individual neurons was determined by employing superparamagnetic clustering (SPC), mean shift clustering algorithm, and K-means clustering algorithm. SPC is an automatic clustering algorithm based on the simulated interaction between each data point and its K nearest neighbors. Here, a range of temperatures (hyperparameter) is required to be pre-specified to automatically determine the number of distinct clusters. Mean shift is an alternative algorithm to the SPC as it automatically selects the number of distinct clusters. However, it has band width as a hyperparameter. Temperature and bandwidth are required to be carefully tuned to achieve the optimal solution to the problem. As SPC and mean shift do not require prior determination of the number of clusters, the threshold of three clusters, which is equal to the real number (ground truth) of clusters is assigned. Once a temperature is fixed in SPC (or a band width is selected in mean shift), if the number of automatically selected clusters is greater than 3, then the 3 largest clusters with most overlapping are considered to calculate the accuracy. In mean shift, a range of bandwidth values are nominated, and the greatest accuracy is reported for the comparison. Lastly, K-means requires a predefined number of clusters for clustering. The number of predefined clusters are estimated and provided by either Silhouette statistics (SH) or gap statistic (GS). The numbers in the parentheses adjacent to values in the k-mean columns indicate the number of clusters, if distinct from 3 (see table 5). We used PCA for feature vectors extraction and CAOM in conjunction with K-means for automatic clustering. Table 5 presents the results of the different combination of the discussed feature extraction algorithms and automatic clustering solutions.

Table 5 shows that PCA/CAOM + K-means, and self-organizing sorter have comprehensively outperformed 7 out of 8 algorithms in terms of mean classification accuracy. However, they provide comparable performance to (DM, SH+K-means). Nonetheless, PCA/CAOM + K-means, and self-organizing sorter perform better than (DM, SH+K-means) in terms of predicting the correct number of neural units on a channel. DM, SH+K-means makes 7 mistakes in predicting the correct number of neural units, however, PCA/CAOM + K-means, and self-organizing sorter made only 2 mistakes as shown in table 5.

PCA/CAOM + K-means and self-organizing sorter provides comparable results in terms of classification accuracy. However, self-organizing sorter has the ability of autonomous fine-tuning which helps to learn the change of data distribution, automatically. Another, major advantage of self-organizing sorter is its suitability of deployment for online spike sorting, which is possible after autonomous fine-tuning. Therefore, self-organizing sorter presents a more desirable solution for spike sorting.

**Table 5**: *Performance comparison of Self-organizing sorter with 9 existing automatic spike classifications algorithms.*

| Examples | Noise levels | No. Spikes | WT SPC | WT Mean Shift | WT GS + K-means | WT GS + K-means | DM SPC | DM Mean shift | DM GS+ K-means | DM SH+ K-means | PCA CAOM + K-means | Convnet Self-organizing sorter |
|---|---|---|---|---|---|---|---|---|---|---|---|---|
| Example 1 | 0.05 | 3514 | 58.56 | 99.22 | 77.96 | 80.72 | 64.48 | 93.33 | 79.69 (5) | 85.66 (5) | 99.23 | **99.71** |
|  | 0.10 | 3522 | 62.68 | 99.16 | 77.24 | 78.56 | 60.48 | 94.17 | 93.14 | 89.60 (4) | 99.31 | **99.59** |
|  | 0.15 | 3477 | 71.19 | 99.20 | 72.81 | 90.03 | 89.14 | 94.14 | 75.43 | 86.12 (4) | **99.13** | **99.18** |
|  | 0.20 | 3474 | 60.52 | 78.89 | 69.57 | 84.80 | 75.60 | 84.93 | 90.58 | **99.42** | 98.93 | 99.01 |
| Example 2 | 0.05 | 3410 | 67.79 | 65.88 | 73.37 | 67.07 | 79.85 | 89.20 | 93.92 | 90.81 (4) | 97.44 | **98.99** |
|  | 0.10 | 3520 | 81.10 | 67.10 | 60.10 | 67.22 | 54.69 | 88.09 | 72.56 | 91.60 (4) | **97.40** | 97.13 |
|  | 0.15 | 3411 | 70.74 | 67.63 | 59.80 | 67.25 | 48.97 | 85.03 | 68.86 (5) | **92.48** | 92.23 | 89.51 |
|  | 0.20 | 3526 | 60.13 | 66.17 | 50.55 | 65.80 | 58.32 | 47.62 | 59.79 (5) | **84.98** | 84.62 | 81.05 |
| Example 3 | 0.05 | 3383 | 63.17 | 34.14 | 73.71 | 65.89 | 66.04 | 87.99 | 73.16 | 87.47 (4) | 97.22 | **98.05** |
|  | 0.10 | 3448 | 53.64 | 33.76 | 51.59 | 61.46 | 36.63 | 65.88 | 88.40 | 92.79 | 93.24 | **95.25** |

|  | 0.15 | 3472 | 66.26 | 33.76 | 47.40 | 41.66 | 33.96 | 33.76 | 88.13 | **86.17** | 67.68(2) | 59(2) |
|  | 0.20 | 3414 | 52.86 | 34.53 | 47.89 | 48.62 | 34.61 | 34.53 | 75.24 | **75.98** | 67.70(2) | 60.05(2) |
| Example 4 | 0.05 | 3364 | 58.69 | 66.51 | 73.23 | 66.62 | 88.88 | 88.92 | 92.93 | 83.95 (5) | 98.57 | **99.19** |
|  | 0.1 | 3462 | 55.36 | 66.98 | 73.42 | 66.96 | 90.70 | 89.52 | 94.09 | 94.22 | 98.65 | **99.26** |
|  | 0.15 | 3440 | 68.11 | 66.66 | 96.96 | 66.63 | 77.03 | 77.83 | 99.01 | **99.01** | 91.86 | 95.04 |
|  | 0.2 | 3493 | 77.75 | 66.94 | 72.36 | 66.93 | 46.02 | 63.10 | 98.68 | **98.68** | 78.83(2) | 73.25(2) |
| Mean accuracy |  |  | 64.28 | 65.41 | 67.37 | 67.89 | 62.83 | 76.12 | 83.97 | **89.94** | **90.75** | **90.20** |

In the result section, we compared BAR and Ada-BAR for the detection of SA in the given dataset. Ada-Bar because of its adaptive nature comprehensively outperformed BAR on the channels where the SA and BA were overlapped.

We used BAR to discard the detected events corresponding to BA. Furthermore, BAR was compared with another method presented in (Quiroga *et al* 2004) on the above-mentioned simulated dataset. In (Quiroga *et al* 2004) spike detection was performed using an automatic amplitude thresholding after bandpass filtering the signal (300–6000 Hz, four pole Butterworth filter). The following setting of threshold (Thr) was used.

$$\text{Thr} = 4\sigma_n; \sigma_n = \text{median}\left\{\frac{|x|}{0.6745}\right\}$$

**Table 6**: *Performance comparison of Ada-BAR with BAR and another method of extracting neural events from the raw data.*

| Examples | Noise levels | No. of Spikes | Misses [Original, BAR, Ada-BAR] | False Positives [Original, BAR, Ada-BAR] |
|---|---|---|---|---|
| Example 1 | 0.05 | 3514 (785) | [17(193), 0 (20), 0(13)] | [711, 190, 145] |
|  | 0.1 | 3522 (769) | [2(177), 2(33), 2(19)] | [57, 11, 17] |
|  | 0.15 | 3477 (784) | [145(215), 43(57), 49(35)] | [14, 5] |
|  | 0.2 | 3474 (796) | [714(275), 118(95), 76(80)] | [10, 2, 2] |
| Example 2 | 0.05 | 3410 (791) | [0(174),0(0), 0(5)] | [0, 0, 0] |

| | 0.1 | 3520 (826) | [0(191), 0(1), 0(7)] | [2, 0, 0] |
| --- | --- | --- | --- | --- |
| | 0.15 | 3411 (763) | [10(173), 14(3), 10(3)] | [1, 0, 0] |
| | 0.2 | 3526 (811) | [376(256), 44(18), 31(21)] | [5, 1, 0] |
| Example 3 | 0.05 | 3383 (767) | [1(210), 3(133), 0(137)] | [63, 18, 9] |
| | 0.1 | 3448 (810) | [0(191), 0(175), 0(150)] | [10, 7, 5] |
| | 0.15 | 3472 (812) | [8(203), 10(190), 5(176)] | [6, 16, 21] |
| | 0.2 | 3414 (790) | [184(219), 170(171), 155(160)] | [2, 5, 9] |
| Example 4 | 0.05 | 3364 (829) | [0(182), 0(2), 0(7)] | [1, 0, 0] |
| | 0.1 | 3462 (720) | [0(152), 0(10), 0(21)] | [5, 3, 0] |
| | 0.15 | 3440 (809) | [3(186), 5(23), 7(29)] | [4, 0, 0] |
| | 0.2 | 3493 (777) | [262(228), 121(41), 90(35)] | [2, 0, 0] |